
\magnification1200
\overfullrule=0pt
\parskip=4pt plus 2pt
\newcount\firstpageno
\firstpageno=2
\footline={\ifnum\pageno<\firstpageno{\hfil}
           \else{\hfil\folio\hfil}\fi}
\def\references{\frenchspacing \parindent=0pt \leftskip=0.8truecm
   \rightskip=0truecm \parskip=4pt plus 2pt \everypar{\hangindent=\parindent}}
\def\endreferences{\vskip0pt}
\def\refstyleprl{
 \gdef\refto##1{~[##1]}				
 \gdef\r##1{~[##1]}	         			
 \gdef\refis##1{\indent\hbox to 0pt{\hss[##1]~}}     	
 \gdef\citerange##1##2##3{~[\cite{##1}--\setbox0=\hbox{\cite{##2}}\cite{##3}]}}
\def\frac#1#2{{\textstyle{#1 \over #2}}} \def\half{{\textstyle{ 1\over 2}}}
  
\def\sss{\scriptscriptstyle}
\def\gtwid{\mathrel{\raise.3ex\hbox{$>$\kern-.75em\lower1ex\hbox{$\sim$}}}}
\def\ltwid{\mathrel{\raise.3ex\hbox{$<$\kern-.75em\lower1ex\hbox{$\sim$}}}}
\refstyleprl
\catcode`@=11
\newcount\r@fcount \r@fcount=0
\newcount\r@fcurr
\immediate\newwrite\reffile
\newif\ifr@ffile\r@ffilefalse
\def\w@rnwrite#1{\ifr@ffile\immediate\write\reffile{#1}\fi\message{#1}}
\def\writer@f#1>>{}
\def\referencefile{
  \r@ffiletrue\immediate\openout\reffile=\jobname.ref%
  \def\writer@f##1>>{\ifr@ffile\immediate\write\reffile%
    {\noexpand\refis{##1} = \csname r@fnum##1\endcsname = %
     \expandafter\expandafter\expandafter\strip@t\expandafter%
     \meaning\csname r@ftext\csname r@fnum##1\endcsname\endcsname}\fi}%
  \def\strip@t##1>>{}}

\def\citeall#1{\xdef#1##1{#1{\noexpand\cite{##1}}}}
\def\cite#1{\each@rg\citer@nge{#1}}	
\def\each@rg#1#2{{\let\thecsname=#1\expandafter\first@rg#2,\end,}}
\def\first@rg#1,{\thecsname{#1}\apply@rg}	
\def\apply@rg#1,{\ifx\end#1\let\next=\relax
\else,\thecsname{#1}\let\next=\apply@rg\fi\next}
\def\citer@nge#1{\citedor@nge#1-\end-}	
\def\citer@ngeat#1\end-{#1}
\def\citedor@nge#1-#2-{\ifx\end#2\r@featspace#1 
  \else\citel@@p{#1}{#2}\citer@ngeat\fi}	
\def\citel@@p#1#2{\ifnum#1>#2{\errmessage{Reference range #1-#2\space is bad.}%
    \errhelp{If you cite a series of references by the notation M-N, then M and
    N must be integers, and N must be greater than or equal to M.}}\else%
 {\count0=#1\count1=#2\advance\count1 by1\relax\expandafter\r@fcite\the\count0,
  \loop\advance\count0 by1\relax
    \ifnum\count0<\count1,\expandafter\r@fcite\the\count0,%
  \repeat}\fi}
\def\r@featspace#1#2 {\r@fcite#1#2,}	
\def\r@fcite#1,{\ifuncit@d{#1}
    \newr@f{#1}%
    \expandafter\gdef\csname r@ftext\number\r@fcount\endcsname%
                     {\message{Reference #1 to be supplied.}%
                      \writer@f#1>>#1 to be supplied.\par}%
 \fi%
 \csname r@fnum#1\endcsname}
\def\ifuncit@d#1{\expandafter\ifx\csname r@fnum#1\endcsname\relax}%
\def\newr@f#1{\global\advance\r@fcount by1%
    \expandafter\xdef\csname r@fnum#1\endcsname{\number\r@fcount}}
\let\r@fis=\refis			
\def\refis#1#2#3\par{\ifuncit@d{#1}
   \newr@f{#1}%
   \w@rnwrite{Reference #1=\number\r@fcount\space is not cited up to now.}\fi%
  \expandafter\gdef\csname r@ftext\csname r@fnum#1\endcsname\endcsname%
  {\writer@f#1>>#2#3\par}}
\def\ignoreuncited{
   \def\refis##1##2##3\par{\ifuncit@d{##1}%
     \else\expandafter\gdef\csname r@ftext\csname r@fnum##1\endcsname\endcsname
     {\writer@f##1>>##2##3\par}\fi}}
\def\r@ferr{\endreferences\errmessage{I was expecting to see
\noexpand\endreferences before now;  I have inserted it here.}}
\let\r@ferences=\references
\def\references{\r@ferences\def\endmode{\r@ferr\par\endgroup}}
\let\endr@ferences=\endreferences
\def\endreferences{\r@fcurr=0
  {\loop\ifnum\r@fcurr<\r@fcount
    \advance\r@fcurr by 1\relax\expandafter\r@fis\expandafter{\number\r@fcurr}%
    \csname r@ftext\number\r@fcurr\endcsname%
  \repeat}\gdef\r@ferr{}\endr@ferences}

\let\r@fend=\endpaper\gdef\endpaper{\ifr@ffile
\immediate\write16{Cross References written on []\jobname.REF.}\fi\r@fend}
\catcode`@=12
\citeall\refto		
\citeall\r		%
\def\la{\bigl\langle} \def\ra{\bigr\rangle} \def\cd{\!\cdot\!} \def\a{\alpha}
\def\b{\beta} \def\g{\gamma} \def\d{\delta} \def\D{\Delta} 
\def\ee{{\rm\sss EE}} \def\p{{\bf p}} \def\P{{\bf P}} \def\Ppm{{\cal P}_\pm}
\def\C{{\cal C}} \def\k{{\bf k}} \def\K{{\bf K}} \def\x{{\bf x}}
\def\X{{\bf X}} \def\s{{\bf s}} \def\S{{\bf S}} \def\y{{\bf y}}
\def\v{{\vphantom{2}}} \def\Na{{\cal N}_\a} \def\Nb{{\cal N}_\b}
\def\zc{Z^\pm_{\rm C}} \def\zgc{Z^\pm_{\rm GC}} \def\Zperp{{\bf Z}_p}
\def\tpsi{{\widetilde\psi}} \def\lam{\lambda} 
\def\ppp{\p_1\ldots\p_N}  \def\paa{\Phi_{\a\a}}
\def\pab{\Phi_{\a\b}} \def\pbb{\Phi_{\b\b}} \def\tpab{\widetilde\Phi_{\a\b}}
\def\fmb{f_{\rm\sss MB}} \def\fbe{f_{\rm\sss BE}} \def\ffd{f_{\rm\sss FD}}
\def\fqm{f_{\rm\sss QM}} \def\intp#1{\int d^3p_{#1} \ldots d^3p_N\,}

 \def\Ubar{{\bar U}}
\def\Tbar{{\bar T}} \def\nbar{{\bar n}} \def\lambar{{\bar\lambda}}
\def\Nc{N_{\rm C}} \def\bc{Berry's conjecture\ }
\def\mb{Maxwell--Boltzmann\ } \def\be{Bose--Einstein\ } \def\fd{Fermi--Dirac\ }
\def\eq{eq.$\,$} \def\eqs{eqs.$\,$}  
\def\fig{fig.$\,$} \def\figs{figs.$\,$}
\baselineskip=13pt
\rightline{cond-mat/9403051}
\rightline{UCSB--TH--94--10}
\rightline{March 1994}
\rightline{Revised April 1994}
\font \bigbf=cmbx12 scaled \magstep2
\vskip 3pt plus 0.3fill
\centerline{\bigbf Chaos and Quantum Thermalization}
\vskip 3pt plus 0.3fill
\font \smcap=cmcsc10
\centerline{\smcap Mark Srednicki}
\vskip 3pt plus 0.1fill
\centerline{\sl Department of Physics\/}
\centerline{\sl University of California\/}
\centerline{\sl Santa Barbara, CA 93106\/}
\centerline{\tt mark@tpau.physics.ucsb.edu}
\vskip 3pt plus 0.3fill
ABSTRACT:~~
We show that a bounded, isolated quantum system of many particles in a specific
initial state will approach thermal equilibrium if the energy eigenfunctions
which are superposed to form that state obey {\it Berry's conjecture}.
Berry's conjecture is expected to hold only if the corresponding classical
system is chaotic, and essentially states that the energy eigenfunctions
behave as if they were gaussian random variables.  We review the existing
evidence, and show that previously neglected effects substantially strengthen
the case for Berry's conjecture.  We study a rarefied hard-sphere gas as an
explicit example of a many-body system which is known to be classically
chaotic, and show that an energy eigenstate which obeys Berry's conjecture
predicts a Maxwell--Boltzmann, Bose--Einstein, or Fermi--Dirac distribution
for the momentum of each constituent particle, depending on whether the wave
functions are taken to be nonsymmetric, completely symmetric, or completely
antisymmetric functions of the positions of the particles.  We call this
phenomenon {\it eigenstate thermalization}.  We show that a generic initial
state will approach thermal equilibrium at least as fast as
$O(\hbar/\Delta)t^{-1}$, where $\Delta$ is the uncertainty in the total
energy of the gas.  This result holds for an individual initial state;
in contrast to the classical theory, no averaging over an ensemble of
initial states is needed.  We argue that these results constitute a
new foundation for quantum statistical mechanics.
\vskip 3pt plus 0.1fill
\centerline{{\sl Phys.~Rev.~E},~~in press}
\vskip 3pt plus 0.3fill \eject
\baselineskip=13pt
\noindent\centerline{\bf I.~~INTRODUCTION}
\vskip0.2in

Take some helium atoms, put them in one corner of a well insulated box, and let
them go.  Wait a while, then punch a small hole in the side of the box.  As the
atoms emerge, one by one, measure their momenta.  Make a histogram, plotting
the fraction of atoms with magnitude of momentum between $p$ and $p+dp$.

Every physicist knows what the result of this experiment will be.
The histogram will be very well approximated by the Maxwell--Boltzmann
distribution
$$\fmb(\p,T) = (2\pi mkT)^{-3/2}\,e^{-\p^2\!/2mkT}                 \eqno(1.1)$$
multiplied by $4\pi F p^2dp$,
where $F=(\pi/8mkT)^{1/2}p$ is a flux factor.
Here $m$ is the mass of a helium atom, $k$ is Boltzmann's constant,
and $T$ is the ``temperature,'' a number which will depend on how the
atoms were originally put into the corner of the box, how much
space they occupied, and other details of the initial conditions.
The challenge is to derive this result from first principles.

The biggest problem with a theoretical analysis of this particular
experiment is the need to treat the hole in the box in a reasonable way.
It is much easier to study the following thought experiment instead.
Suppose, after preparing the system in its initial state, we are
able to measure the momentum of one atom at a specific (but
arbitrary) time $t$.  Suppose further that, after having made this
measurement, we can empty out the box, and then start it off again
with the system in exactly the same initial state.
We now do this repeatedly, each time measuring the momentum of one atom
after exactly the same amount of time~$t$ has passed.
We make a histogram of the results.

Let us analyze this experiment, beginning with classical mechanics
as the underlying theory.  We take the hamiltonian for $N$ atoms to be
$$H = \sum_{i=1}^N {\p_i^2\over2m} + \sum_{i<j}V\bigl(|\x_i-\x_j|\bigr)\;,
                                                                 \eqno(1.2)$$
where we take $V(r)$ to be a hard-sphere potential:
$$V(r) = \cases{+\infty & for $r<2a$ \cr
\noalign{\medskip}
                      0 & for $r>2a\;.$ \cr}                   \eqno(1.3)$$
We assume perfectly reflecting boundary conditions at the walls of the box.
The atoms initially have some definite total energy $U$.  The phase space of
this system is known to be fully chaotic, with no invariant tori for any value
of $U$\r{sinai}.  Thus the motion in phase space on any constant energy
surface is ergodic and mixing.  (For a review of classical chaos theory,
see, e.g.,\citerange{berry79}{zav}{gutz}.)

However, this is entirely irrelevant if we always start out with exactly
the same initial state, and always make the measurement after exactly the same
amount of time has elapsed.  The momentum of the measured atom (assuming
that it is always the same atom) is determined exactly by the initial
conditions, and so will always be the same.  To have any hope of getting
a distribution of momenta, we must average over either the initial
conditions or the times of measurement or both.

If we keep the initial conditions fixed, ergodicity implies that the system
wanders all over the constant energy surface.  (This assumes that we have not
started the system off at a point located on a periodic orbit; such points
form a set of measure zero.) If we divide the constant energy surface into
many patches of equal area (and that area is not too small) then after a
certain finite time the system will, to a very good approximation,
be equally likely to be in any one of these patches at any later time.
Conversely, if we permit a range of initial conditions, mixing implies
that, if the measurement time is fixed but not too early, the system
will once again, to a very good approximation, be equally likely to be
in any one of the equal-area patches.
The rule that equal-area patches are equally likely is just the usual
formulation of the microcanonical ensemble, which after a little work leads
to \eq(1.1) for the fraction of atoms with momentum in a range of $d^3p$
around $\p$, with the temperature $T$ simply given by the ideal gas formula
$U=\frac32 NkT$.  In short, if we do a modest average over either the initial
conditions or the times of measurement, then classical chaos results in
classical thermalization.  (For an elementary review of chaos theory as it
applies to classical statistical mechanics, see\r{reichl}.)

On the other hand, if we have a weakly perturbed integrable system
(for example, harmonic oscillators with small nonlinear couplings),
then according to the Kolmogorov--Arnol'd--Moser theorem\r{KAM},
its phase space is foliated by invariant tori almost everywhere,
and we do not expect it to thermalize.  If it is partially integrable,
with some invariant tori embedded in an otherwise chaotic phase space,
then the system may or may not thermalize, depending on the initial state.

These results from classical mechanics are clear and powerful, and provide a
satisfying explanation of statistical behavior in classical systems which
exhibit chaos.  However, we know that the real world is ultimately described by
quantum mechanics, and we so should seek the quantum analog of the classical
analysis.  We would like to know, for example, what property a quantum system
must possess, analogous to classical chaos, so that ``most'' of its initial
states thermalize, in the sense discussed above.  Furthermore if a quantum
system does possess this property (whatever it may be), then we might hope that
the inherent uncertainties in quantum mechanics lead to a thermal distribution
for the momentum of a single atom, even if we always start with exactly the
same initial state, and make the measurement at exactly the same time.  If this
is true, then quantum mechanics automatically provides the ``coarse graining''
which is missing\r{zav,krylov} in the classical theory.

I will argue that the property needed for thermalization of a quantum system
is the validity of
{\it Berry's conjecture}\citerange{berry77b}{berry81}{berry91}.
For a quantum gas of hard spheres, \bc states that each energy eigenfunction
appears to be a superposition of plane waves (in the $3N$ dimensional
coordinate space) with random phase and gaussian random amplitude, but fixed
wavelength.  In general, \bc is expected to hold only for systems which
exhibit classical chaos in all or at least most of the classical phase space.
As already noted, a hard sphere gas meets this condition.

We will see that Berry's conjecture leads to either Bose--Einstein,
Fermi--Dirac, or Maxwell--Boltzmann statistics, depending on whether the
wave functions are chosen to be completely symmetric, completely antisymmetric,
or nonsymmetric functions of the positions of the $N$ atoms.
Furthermore we will find that any nonthermal features
of the initial distribution of momenta decay away at least as fast as
$O(h/\D)t^{-1}$, where $h$ is Planck's constant and
$\D$ is the uncertainty in the total energy.
Thermal behavior thus appears for a very wide range of possible initial states,
without assuming that the system interacts with an external heat bath, or any
other environmental variables.  We also do not need to take any averages over
initial states, times of measurement, or hamiltonians, or make any
unjustifiable approximations to the quantum equation of motion,
such as truncation of the BBGKY hierarchy.

The rest of this paper is organized as follows.
In Sect.$\;$II,
we review \bc for a system of hard spheres in a box.
In Sect.$\;$III,
we show how Berry's conjecture leads, in the limit of low density and high
energy, to a Maxwell--Boltzmann distribution for the momentum of a single atom
in the gas, with a temperature that is related
to the total energy by the ideal gas law.
In this section we treat the atoms as
distinguishable, making no assumptions about the symmetry of the wave
function of the gas under exchange of individual atoms.
In Sect.$\;$IV,
we evaluate the effects of certain corrections to \bc known as ``scars,''
and reconsider some of the numerical results on \bc which have appeared
in the literature.
In Sect.$\;$V,
we examine time evolution beginning with a nonthermal initial state,
and study the approach to equilibrium.
In Sect.$\;$VI,
we consider wave functions which are completely symmetric or
antisymmetric under exchange of atoms, and see that these lead to \be or \fd
distributions for individual momenta, respectively.
Conclusions, speculations, and possible extensions are presented
in Sect.$\;$VII.

\vskip0.2in
\noindent\centerline{\bf II.~~BERRY'S CONJECTURE}
\vskip0.2in

Consider a system of $N$ hard spheres, each of radius $a$, in a cubic box with
edge length $L+2a$.  Call the energy eigenvalues $U_\a$ and the corresponding
eigenfunctions $\psi_\a(\X)$, where $\X=(\x_1,\ldots,\x_N)$ denotes
the $3N$ coordinates, and $\P=(\p_1,\ldots,\p_N)$ will denote the
$3N$ conjugate momenta.  We take the wave functions to be defined on the domain
$$D=\Bigl\{~\x_1,\ldots,\x_N~\Big|~-\half L\le x_{i1,2,3}\le
                +\half L;~|\x_i-\x_j|\ge 2a~\Bigr\}\;,    \eqno(2.1)$$
with the boundary condition that each $\psi_\a(\X)$ vanish on the boundary
of $D$.  For now we assume that $\psi_\a(\X)$
has no symmetries under exchange of individual $\x_i$.

The energy eigenfunctions $\psi_\a(\X)$ can always be chosen to be
everywhere real, and can be written as
$$\psi_\a(\X) = \Na\int d^{3N}\!\!P\;A_\a(\P)\;
                   \d\bigl(\P^2-2m U_\a\bigr)\,
                   \exp(i\P\cd\X/\hbar) \;,                     \eqno(2.2)$$
where $\Na$ is a constant to be determined by the normalization condition
$$\int_D d^{3N}\!\!X\;\psi^2_\a(\X) = 1\;,                      \eqno(2.3)$$
and where $A_\a^*(\P)=A_\a(-\P)$.  For this system, \bc is equivalent to
assuming that $A_\a(\P)$ can be treated as a gaussian random variable with a
two-point correlation function given by
$$\la A_\a(\P)A_\b(\P')\ra_\ee
               = \d_{\a\b}\,\d^{3N}(\P+\P')/\d(\P^2-\P'^2)\;.  \eqno(2.4)$$
Here $\d^{3N}(\P)$ is the $3N$-dimensional Dirac delta function
and $\d(x)$ is the one-dimensional Dirac delta function.
The subscript EE stands for ``eigenstate ensemble.''  This is a
fictitious ensemble which describes the properties of a typical
energy eigenfunction.  Individual eigenfunctions behave {\it as if\/} they
were selected at random from the eigenstate ensemble.
Berry's conjecture also asserts that the eigenstate ensemble is gaussian,
so that all multipoint correlation functions are given in terms of the
two-point correlation function; e.g.,
$$\eqalignno{
    \la A_\a(\P_1)A_\b(\P_2)               A_\g(\P_3)A_\d(\P_4)\ra_\ee &=
    \la A_\a(\P_1)A_\b(\P_2)\ra_\ee \; \la A_\g(\P_3)A_\d(\P_4)\ra_\ee \cr
\noalign{\smallskip}
&\,+\la A_\a(\P_1)A_\g(\P_3)\ra_\ee \; \la A_\d(\P_4)A_\b(\P_2)\ra_\ee \cr
\noalign{\smallskip}
&\,+\la A_\a(\P_1)A_\d(\P_4)\ra_\ee \; \la A_\b(\P_2)A_\g(\P_3)\ra_\ee \;.
                                                                   &(2.5)\cr}$$
Of course, each $A_\a(\P)$ must give back a $\psi_\a(\X)$ which vanishes
on the boundary of $D$; this is not a stringent requirement on $A_\a(\P)$
at high energy, where $\psi_\a(\X)$ has many wavelengths between any two
segments of the boundary.  We will say more about the requirement of high
energy shortly.  Meanwhile, for a more general but less transparent definition
of Berry's conjecture, see sect.$\;$4; for related mathematical results,
see\r{voros,math}.

Berry's conjecture is based on semiclassical reasoning, and is manifestly
untrue for systems whose classical phase space is foliated almost everywhere by
invariant tori\r{berry77a}.  It has been investigated numerically for simple
systems which are fully chaotic classically, such as a single particle in a
two-dimensional stadium-shaped box\citerange{mackauf79}{heller84,mackauf88,
shapiro88}{heller91}, or a single particle on a two-dimensional surface with
constant negative curvature and periodic boundary conditions\r{aurstein91,
aurstein93}.  In these systems, \bc is found to be valid for eigenstates of
sufficiently high energy, and its validity
has even been suggested as a good {\it definition\/} of chaos in a quantum
system\r{steiner94}.  However, an important caveat is the existence of
``scars'' on some energy eigenfunctions, regions of enhanced value of
$\psi_\a(\X)$ which follow the paths of the most stable classical periodic
orbits\r{mackauf79,heller84}.  For now we will ignore the scars, since their
presence will not alter any of our conclusions.  We will discuss them in more
detail in sect.$\;$4.  Also, there we will argue that some numerical results
which have been interpreted as evidence against \bc actually provide evidence
for it.

Even in a system which is fully chaotic classically (like a hard-sphere gas),
\bc will certainly not be valid for eigenfunctions which are too low in energy.
The low-lying states necessarily have specific structure:  the ground state,
for example, is nodeless.  A rough criterion for the validity of \bc is that
the average wavelength of each particle be small enough to ``see'' the features
which produce classical chaos\r{mackauf88}.
For the hard-sphere gas, the relevant feature is the nonzero radius $a$ of
each particle.  Anticipating a bit and defining a temperature $T_\a$ for
each energy eigenvalue $U_\a$ via the ideal gas formula $U_\a=\frac32 NkT_\a$,
and further defining a ``thermal'' wavelength
$\lam_\a=(2\pi\hbar^2/mkT_\a)^{1/2}$, then the criterion for the validity
of \bc is $\lam_\a\ltwid a$.  Numerically, this becomes
$T_\a\gtwid(300/ma^2)\;$Kelvin, where $a$ is in angstroms and $m$ is in amu.

It turns out that getting explicit results will also require us to work at low
density, $Na^3\ll L^3$.  Combining this with $\lam_\a\ltwid a$, we see that we
need to have $N\lam_\a^3\ll L^3$, a condition which is also required, in
quantum statistical mechanics, for the \be and \fd distributions to be well
approximated by the \mb distribution.

Let us now consider the eigenfunctions in momentum space:
$$\eqalignno{\tpsi_\a(\P)
   &\equiv h^{-3N/2}\int_D d^{3N}\!\!X\;\psi_\a(\X)\,\exp(-i\P\cd\X/\hbar)\cr
\noalign{\medskip}
   &=h^{3N/2}\;\Na\int_{-\infty}^{+\infty}d^{3N}\!\!K\;A_\a(\K)\,
     \d\bigl(\K^2-2mU_\a\bigr)\,\d^{3N}_D(\K-\P) \;,             &(2.6)\cr}$$
where we have defined
$$\d^{3N}_D(\K) \equiv h^{-3N}\int_D d^{3N}\!\!X\;\exp(i\K\cd\X/\hbar)\;.
                                                                \eqno(2.7)$$
If the condition needed for \bc is satisfied ($\lam_\a\ltwid a$),
{\it and\/} we are in the low density regime ($Na^3\ll L^3$),
then we can make the substitutions
$$\eqalign{
\d^{3N}_D(0) &\to (L/h)^{3N}\;,\cr
\noalign{\medskip}
\d^{3N}_D(\P) &\to \d^{3N}(\P)\;,\cr
\noalign{\medskip}
\bigl[\d^{3N}_D(\P)\bigr]{}^2 &\to (L/h)^{3N}\d^{3N}(\P)\;. \cr}  \eqno(2.8)$$
Now using \eqs(2.2), (2.4), (2.6), and (2.8), we find that
(for $\lam_\a\ltwid a$ and $Na^3\ll L^3$),
$$\la\tpsi^*_\a(\P)\tpsi_\b(\P')\ra_\ee = \d_{\a\b}\,\Na^2\,h^{3N}\,
                \d(\P^2-2mU_\a)\,\d^{3N}_D(\P-\P')\;,             \eqno(2.9)$$
which will play a key role in the next section.

\vskip0.2in
\noindent\centerline{\bf III.~~EIGENSTATE THERMALIZATION}
\vskip0.2in

Let us now put our gas of $N$ hard spheres into some initial state specified
by the momentum-space wave function
$$\tpsi(\P,0)=\sum_\a C_\a\,\tpsi_\a(\P)\;.                       \eqno(3.1)$$
We take the energy eigenfunctions to be orthonormal, and also assume
$\tpsi(\P,0)$ to be normalized, so that $\sum_\a |C_\a|^2=1$.
The expectation value of the energy is then
$$\Ubar = \sum_\a |C_\a|^2\,U_\a\;,                               \eqno(3.2)$$
and the uncertainty in the energy is $\D$, where
$$\D^2 = \sum_\a |C_\a|^2\,\bigl(U_\a-\Ubar\bigr)^2\;.            \eqno(3.3)$$
We will assume that $\D\ll\Ubar$.
The initial wave function will evolve in time according to the
Schrodinger equation:
$$\tpsi(\P,t) = \sum_\a C_\a\exp(-iU_\a t/\hbar)\,\tpsi_\a(\P)\;.  \eqno(3.4)$$
Now return to the thought experiment
in which the system is repeatedly prepared in the same initial state
(specified by the $C_\a$'s), and the momentum of one atom is repeatedly
measured after the same elapsed time~$t$.  The theoretical prediction
for the fraction of atoms with momentum in a range $d^3p$ around $\p$ is
$\fqm(\p,t)d^3p$, where
$$\eqalignno{
\fqm(\p_1,t) &= \intp2 \bigl|\tpsi(\P,t)\bigr|{}^2 \cr
\noalign{\medskip}
             &= \sum_{\a\b}C_\a^* C_\b\,e^{i(U_\a-U_\b)t/\hbar}
                \intp2 \tpsi^*_\a(\P)\tpsi_\b(\P)\cr
\noalign{\medskip}
             &= \sum_{\a\b}C_\a^* C_\b\,e^{i(U_\a-U_\b)t/\hbar}
                \;\pab(\p_1)\;.                                   &(3.5)\cr}$$
In the last line we have introduced
$$\Phi_{\a\b}(\p_1)\equiv\intp2\tpsi_\a^*(\P)\tpsi_\b(\P)\;,\eqno(3.6)$$
which obeys the normalization condition
$$\int d^3p_1\,\Phi_{\a\b}(\p_1)=\d_{\a\b}\;.               \eqno(3.7)$$
If the system thermalizes, then after some time has passed,
$\fqm(\p_1,t)$ should be equal to the \mb distribution of \eq(1.1),
although (as in the classical case) some modest averaging over either the
initial conditions (the $C_\a$'s) or the times of measurement
(the value of $t$) might first be necessary.  Furthermore the temperature
$\Tbar$ should be given at least approximately by the ideal gas law
$\Ubar=\frac32Nk\Tbar$, with a fractional uncertainty of order $\D/\Ubar$.

To practice on a simple example, let us study the case where the
initial state is a single energy eigenstate.  This is, of course, unphysical:
we cannot actually prepare such a state in a time less than $O(\hbar/\d)$,
where $\d$ is the mean energy level spacing near $U_\a$\r{landau}.
This is fantastically small in any realistic case [$\d_\a = 1/n_\a$, where
$n_\a$ is given by \eq(4.6)], so that $\hbar/\d_\a$ is much longer than the age
of the universe. Nevertheless taking the initial state to be an energy
eigenstate will turn out to be an instructive exercise.

In this case, \eq(3.5) becomes simply $\fqm(\p_1,t)=\paa(\p_1)$, which is
independent of time.  We now study the properties of $\paa(\p_1)$ in
the eigenstate ensemble introduced in the previous section.  Assuming
high energy ($\lam_\a\ltwid a$) and low density ($Na^3\ll L^3$),
it follows from \eqs(2.8), (2.9), and (3.6) that
$$\la\paa(\p_1)\ra_\ee=\Na^2\,L^{3N}\intp2 \d(\P^2-2mU_\a)\;.
                                                                  \eqno(3.8)$$
We introduce the useful formula
$$\eqalignno{
I_D(x) &\equiv \int d^D\!P\;\d({\bf P}^2-x) \cr
\noalign{\medskip}
       &= {(\pi x)^{D/2}\over\Gamma(D/2)x}\;,                     &(3.9)\cr}$$
and use it to fix $\Na^{-2}=L^{3N}I_{3N}(2mU_\a)$ via \eq(3.7).
We then find
$$\eqalignno{
\la\paa(\p_1)\ra_\ee &= {I_{3N-3}(2mU_\a-\p_1^2)\over I_{3N}(2mU_\a)} \cr
\noalign{\medskip}
                     &= {\Gamma(3N/2)\over\Gamma((3N-3)/2)}
                        \left({1\over2\pi mU_\a}\right)^{\!3/2}
                        \biggl(1-{\p_1^2\over2mU_\a}\biggr)^{\!(3N-5)/2}\;.
                                                                 &(3.10)\cr}$$
If we now set $U_\a=\frac32NkT_\a$ and take the large $N$ limit, we get
$$\la\paa(\p_1)\ra_\ee = (2\pi mkT_\a)^{-3/2}\,e^{-\p_1^2/2mkT_\a}\;,
                                                                 \eqno(3.11)$$
which is precisely $\fmb(\p_1,T_\a)$.

Let us note first that, given \eq(3.8) as a starting point, \eqs(3.9--11)
simply recapitulate a standard derivation of the canonical ensemble from the
microcanonical\r{jancel}.

More importantly, we must study the fluctuations of $\paa(\p_1)$ about
its average value in the eigenstate ensemble.  We begin by defining
$$\bigl[\D\pab(\p_1)\bigr]{}^2 \equiv
                          \la|\pab(\p_1)|^2\ra_\ee
                       - \bigl|\la\pab(\p_1)\ra_\ee\bigr|{}^2 \;. \eqno(3.12)$$
Using \eqs(2.5), (2.8), (2.9), and (3.6), we find
$$\eqalignno{
\bigl[\D\pab(\p_1)\bigr]{}^2
      &= \Na^2\Nb^2\,(Lh)^{3N}\intp2 d^3p'_2\ldots d^3p'_N\cr
      &\qquad\times\d(\P^2-2mU_\a)\,\d(\P'^2-2mU_\b)\,\d^{3N}_D(\P-\P')\;.
                                                                 &(3.13)\cr}$$
Before evaluating \eq(3.13) explicitly, we can see that it will be very small:
if we replace $\d^{3N}_D(\P-\P')$ by its maximum value $(L/h)^{3N}$ everywhere,
the right-hand side of \eq(3.13) becomes
$\la\paa(\p_1)\ra_\ee\,\la\pbb(\p_1)\ra_\ee$ [cf.~\eq(3.8)].
Of course this replacement results in a huge overestimate of
$\bigl[\D\pab(\p_1)\bigr]{}^2$, since in fact $\d^{3N}_D(\P-\P')$ is close to
zero almost everywhere.  Thus we will have, in particular,
$\D\paa(\p_1)\ll\la\paa(\p_1)\ra_\ee$.  Furthermore we see why
$\D\paa(\p_1)$ is so small: $\tpsi_\a(\P)$ has fluctuations of order one in the
eigenstate ensemble, but these are washed out when we integrate over most of
the momenta.  (In Sect.$\;$IV we will see that the same fate befalls the
``scars'' mentioned in Sect.$\;$II.)

We now turn to the evaluation of \eq(3.13), to find out just how small it is.
We will need to know a bit more detail about $\d^{3N}_D(\P-\P')$ than the
substitution rules of \eq(2.8).  We therefore approximate it with a gaussian:
$$\d^{3N}_D(\P-\P') \simeq (L/h)^{3N}
                    \exp\bigl[-(\P-\P')^2L^2/4\pi\hbar^2\bigr]\;.\eqno(3.14)$$
In the low density regime ($Na^3\ll L^3$), using \eq(3.14) instead of
\eq(2.7) changes the result only by an overall constant of order one.
Substituting \eq(3.14) into \eq(3.13), setting $\a=\b$,
and taking the large $N$ limit yields
$$\D\paa(\p_1) = O(1)\,N^{1/2}\,e^{-3N/4}\,(L/\lam_\a)^{-(3N-6)/2}\,
                 e^{+\p_1^2\!/4mkT_\a}\fmb(\p_1,T_\a) \;.         \eqno(3.15)$$
Since we have $L\gg\lam_\a$, we see that
{\it the fluctuations in $\paa(\p_1)$ about $\fmb(\p_1,T_\a)$ are
negligibly small for large $N$.}  That is to say, an energy eigenstate
which satisfies Berry's conjecture predicts a thermal distribution
for the momentum of a single constituent particle.
We will refer to this remarkable phenomenon as {\it eigenstate thermalization}.

Given a system which exhibits eigenstate thermalization, it is not hard to
understand why almost any initial state will thermalize.  In fact, the problem
now is primarily to {\it prevent\/} the system from having a thermal
distribution for the momentum of each particle at {\it all\/} times.  To do so
at $t=0$, we must carefully superpose energy eigenstates in order to construct
an initial state with whatever nonthermal features we might want.  Once this
superposition is prepared, however, the delicate phase relationships we have
set up to avoid thermal behavior will gradually be destroyed by hamiltonian
time evolution, and the system will equilibrate.  We will see how this works in
more detail in Sect.$\;$V.  First, however, we digress briefly to discuss the
scars, and check to see that they do not change any of our conclusions so far.

\vskip0.2in
\noindent\centerline{\bf IV.~~FADED SCARS}
\vskip0.2in

The theory of scar formation has been developed by Heller\r{heller84,heller91},
Bogomolny\r{bog}, and Berry\r{berry89,berry91}.
We will be rather schematic here; readers unfamiliar with scar theory
should consult the cited references for more details.

We first consider any system governed by a hamiltonian $H(\P,\X)$ which is
time-reversal invariant and which results in classical chaos.  For consistency
of notation with the previous sections,
we take $\P$ and $\X$ to be vectors with $3N$ components.

We begin by introducing the Wigner density for an eigenstate,
$$\rho_\a(\P,\X) = h^{-3N}\int d^{3N}\!\!S\;\exp(i\P\cdot\S/\hbar)\;
                   \psi_\a\bigl(X+\half\S\bigr)
                   \psi_\a\bigl(X-\half\S\bigr)\;,                 \eqno(4.1)$$
where $\psi_\a(\X)$ is real.  The Wigner density has the useful properties that
$$\eqalign{ \int d^{3N}\!\!P\;\rho_\a(\P,\X) &= \psi^2_\a(\X)\;,\cr
\noalign{\medskip}
\int d^{3N}\!\!X\;\rho_\a(\P,\X)
                              &=\bigl|\tpsi_\a(\P)\bigr|{}^2\;,\cr}\eqno(4.2)$$
which imply the normalization condition
$$\int d^{3N}\!\!P\;d^{3N}\!\!X\;\rho_\a(\P,\X) = 1\;.             \eqno(4.3)$$
Scar theory begins with a semiclassical formula for $\rho_\a(\P,\X)$:
$$\rho_\a(\P,\X) = n_\a^{-1}\,h^{-3N}\,\d\bigl(H(\P,\X)-U_\a\bigr)\,
                   \Bigl\{ 1 + \sum_p A_p\,e^{iS_p/\hbar}
                           \exp\bigl[(i/\hbar)\Zperp\cd{\bf W}_p(T_p)\cd\Zperp
                                     \bigr]\Bigr\} \;.             \eqno(4.4)$$
The sum is over all periodic orbits on the surface with constant
energy $U_\a$; $S_p$ is the action of the orbit; $T_p$ is a coordinate in
phase space along the orbit; $\Zperp$ are the $6N-2$ coordinates
in the energy surface which are perpendicular to the orbit;
and $A_p$ and ${\bf W}_p(T_p)$ are purely
classical quantities which depend on the monodromy matrix of the orbit.
The constant $n_\a$ is fixed by the normalization condition, \eq(4.3), and
can be interpreted physically as the energy eigenvalue density near $U_\a$.
If we ignore the sum over periodic orbits in \eq(4.5), we obtain the
``Weyl rule" for $n_\a$:
$$n_\a = h^{-3N}\int d^{3N}\!\!P\;d^{3N}\!\!X\;
         \d\bigl(H(\P,\X)-U_\a\bigr)\;.                            \eqno(4.5)$$
In the case of a hard-sphere gas of $3N$ distinguishable particles,
this becomes
$$n_\a = {1 \over \Gamma(3N/2)U_\a}
         \biggl({mL^2 U_\a \over 2\pi\hbar^2}\biggr)^{\!3N/2}\;.   \eqno(4.6)$$
For bosons or fermions, the right-hand side should be divided by $N!$.
Even so, $n_\a$ is fantastically large in any realistic case\r{landau}.

The key point for scar theory is that the periodic-orbit terms in \eq(4.4)
have no $\hbar$-dependent prefactors; the peak height of each term is
controlled by the classical quantity $A_p$.  A short periodic orbit can have
an $A_p$ which is greater than one; this produces an obvious ``scar" in phase
space along the path of the orbit.

We are interested, however, in $\paa(\p_1)$, and so we must integrate
$\rho_\a(\P,\X)$ over all $3N$ components of $\X$, and all but three of the
$3N$ components of $\P$.  If we consider isolated periodic orbits, we see
from \eq(4.4) that an integral over one of the $6N-2$ components of $\Zperp$
yields a prefactor of $\hbar^{1/2}$.  Thus the contribution of each isolated
periodic orbit to $\paa(\p_1)$ is suppressed, relative to the leading term, by
$\hbar^{(6N-3)/2}$, which means that each individual scar on $\paa(\p_1)$ is
totally negligible.

Nonisolated periodic orbits are a little more complicated, since moving off a
nonisolated orbit in some directions in phase space merely puts the system onto
another nonisolated orbit in the same ``family''; there are a finite number
of these families.  For the hard-sphere gas, the nonisolated orbits consist of
motions where the spheres bounce off the walls but never collide with each
other\r{gaspard}.
A given nonisolated orbit of this type can
in general be deformed into another one by moving in any direction in
coordinate space.  Thus when integrating $\rho_\a(\P,\X)$
over the $3N$ components of $\X$, we do not get any factors of $\hbar^{1/2}$.
We do, however, get a net factor of $\hbar^{(3N-3)/2} $ from integrating
over $3N-3$ components of $\P$.  Thus, while the contribution of a family of
nonisolated orbits to $\paa(\p_1)$ is much larger than the contribution of a
single isolated orbit, it is still negligible.  These conclusions are
supported by the much more detailed calculation of Gaspard\r{gaspard} for
the periodic-orbit corrections to $n_\a$ for the hard-sphere gas.

Therefore, in computing $\paa(\p_1)$, we can safely ignore the short
isolated orbits and all of the nonisolated orbits.  The most modern version of
Berry's conjecture\r{berry91} then assigns the origin of the gaussian
fluctuations in $\psi_\a(\X)$ to the long isolated orbits.  Thus $\psi_\a(\X)$
is conjectured to behave like a gaussian random variable with a two-point
correlation function embodied by the elegant formula\r{berry77b,voros}
$$\la\rho_\a(\P,\X)\ra_\ee = n_\a^{-1}\,h^{-3N}\,\d\bigl(H(\P,\X)-U_\a\bigr)\;.
                                                                   \eqno(4.7)$$
There has been important progress recently\r{berrykeat} in bringing
the long isolated orbits under analytic control, but so far \bc remains just
that.  Even if a rigorous proof is eventually found, it is likely to apply
only for asymptotically high energies.  We turn, therefore, to a discussion
of the existing numerical evidence.

\bc has been studied numerically in some two-dimensional systems, such as a
particle in a stadium-shaped box\citerange{mackauf79}{heller84,mackauf88,
shapiro88}{heller91}.  One popular object to study is the correlation
function\r{berry77b}
$$\eqalignno{
\C_\a(\s) &\equiv \int d^2x\;\psi_\a\bigl(\x+\half\s\bigr)
                         \psi_\a\bigl(\x-\half\s\bigr)\cr
\noalign{\medskip}
     &=\int d^2p\;d^2x\;\exp(-i\p\cd\s/\hbar)\,\rho_\a(\p,\x)\;,   &(4.8)\cr}$$
where in the simplest case the integral over $x$ covers the entire box, whose
area we will call $L^2$.  For a particular eigenstate, the numerically computed
$\C_\a(\s)$ is compared to its expectation value in the eigenstate ensemble:
$$\eqalignno{
\la \C_\a(\s)\ra_\ee &= \int d^2p\;d^2x\;\exp(-i\p\cd\s/\hbar)\,
                                     \la\rho_\a(\p,\x)\ra_\ee\cr
\noalign{\smallskip}
                     &= n_\a^{-1}\,h^{-2}\,L^2 \int d^2p\;\d(\p^2\!/2m-U_\a)\cr
\noalign{\smallskip}
                     &= J_0(k_\a s) \;,                           & (4.9)\cr}$$
where $J_0(z)$ is a Bessel function, $k_\a=(2mU_\a/\hbar^2)^{1/2}$, and
$s=|\s|$.  In\r{mackauf88}, only moderately good agreement was found with this
prediction, with discrepancies of approximately 0.1 for $k_\a=65$ and
$L=\sqrt\pi$; see \fig7 of\r{mackauf88}.  However, these discrepancies
are entirely explained by consideration of the fluctuations in $\C_\a(\s)$
which are predicted by the eigenstate ensemble:
$$\eqalignno{
\bigl[\D\C_\a(\s)\bigr]{}^2 &\equiv \la\C^2_\a(\s)\ra_\ee
                                  - \la\C_\a(\s)\ra^2_\ee \cr
\noalign{\medskip}
                            &= \int d^2x\;d^2y\,\Bigl[
                               \la\psi_\a\bigl(\x+\half\s\bigr)
                                  \psi_\a\bigl(\y+\half\s\bigr)\ra_\ee\,
                               \la\psi_\a\bigl(\x-\half\s\bigr)
                                  \psi_\a\bigl(\y-\half\s\bigr)\ra_\ee \cr
\noalign{\smallskip}
                            &\qquad\qquad\quad\; +
                               \la\psi_\a\bigl(\x+\half\s\bigr)
                                  \psi_\a\bigl(\y-\half\s\bigr)\ra_\ee\,
                               \la\psi_\a\bigl(\x+\half\s\bigr)
                                  \psi_\a\bigl(\y-\half\s\bigr)\ra_\ee\Bigr]\cr
\noalign{\medskip}
                            &= L^{-4} \int d^2x\;d^2y\,\Bigl[
                                           J_0^2\bigl(k_\a|\x-\y|\bigr)
                                         + J_0\bigl(k_\a|\x-\y+\s|\bigr)
                                           J_0\bigl(k_\a|\x-\y-\s|\bigr)
                                      \Bigr]\;. \cr
                            & {}                                 & (4.10)\cr}$$
The first term in the last line dominates over the second for all $\s$,
and is $O(1/k_\a L)$.  That is, we expect discrepancies of roughly
$(k_\a L)^{-1/2}$ between $\C_\a(\s)$ as computed numerically for a
particular eigenstate and $\la\C_\a(\s)\ra_\ee$ as given by \eq(4.9).
This is {\it exactly what is seen\/} in \fig7 of\r{mackauf88}.
Similar comments apply to \figs14--17 of\r{aurstein93}.
The fact that these discrepancies are {\it predicted\/}
by the eigenstate ensemble does not seem to have been noticed previously.

We note finally that the gaussian nature of the eigenstate ensemble [which is
used crucially in \eq(4.10)] has also been directly tested.
The prediction is that
$$\la\psi_\a^{2n}(\x)\ra_\ee = (2n-1)!!\,\la\psi_\a^2(\x)\ra^n_\ee
                             = (2n-1)!!\,L^{-2n} \;,              \eqno(4.11)$$
but this just means that the probability distribution for the amplitude $\psi$
at any point is
$$P(\psi)={L\over\sqrt{2\pi}}\exp\bigl[-\half L^2\psi^2\bigr]\;,  \eqno(4.12)$$
which is well supported by the numerical
results\r{mackauf88,aurstein91,aurstein93}.

\vskip0.2in
\noindent\centerline{\bf V.~~TIME EVOLUTION AND EQUILIBRATION}
\vskip0.2in

In Sect.$\;$III, we saw that an individual energy eigenstate which satisfies
\bc predicts a thermal distribution for the momentum of each constituent
particle.  Now we must see what happens when we consider more general states.
We will once again express the initial state as a wave function in momentum
space, and expand it in energy eigenfunctions, as in \eq(3.1).  We assume that
the uncertainty $\D$ in the total energy, \eq(3.3), is much smaller than the
average energy $\Ubar$, \eq(3.2).  This is easy to arrange in practice.

The predicted momentum distribution of a single particle at time $t$
is $\fqm(\p_1,t)$, as given by \eq(3.5).  Now take the average of \eq(3.5)
in the eigenstate ensemble.  From \eq(2.4), it is immediately obvious
that $\la\pab(\p_1)\ra_\ee=0$ if $\a\ne\b$, and so we get
$$\eqalignno{
  \la\fqm(\p_1,t)\ra_\ee &= \sum_\a |C_\a|^2\,\la\paa(\p_1)\ra_\ee \cr
\noalign{\medskip}
                         &= \sum_\a |C_\a|^2\,
                   (2\pi mkT_\a)^{-3/2}\,e^{-\p_1^2/2mkT_\a}\;,  &(5.1)\cr}$$
where we have used \eq(3.11) in the second line.
Now, since $\D\ll\Ubar$, we can with negligible error replace each $T_\a$
in this sum with $\Tbar$, where $\Ubar=\frac32 Nk\Tbar$, and then
$\sum_\a|C_\a|^2=1$ gives us
$$\la\fqm(\p_1,t)\ra_\ee=\fmb\bigl(\p_1,\Tbar\bigr)
                        \bigl[1+O\bigl(\D/\Ubar\bigr)\bigr]\;,   \eqno(5.2)$$
the desired result.

Once again, though, we must study the fluctuations
of $\fqm(\p_1,t)$ that are predicted by the eigenstate ensemble.
We can write
$$\fqm(\p_1,t)=\fmb\bigl(\p_1,\Tbar\bigr)
              +\sum_{\a\b}C^*_\a C_\b\,e^{i(U_\a-U_\b)t/\hbar}\,\tpab(\p_1)\;,
                                                                 \eqno(5.3)$$
where we have defined
$$\tpab(p_1) \equiv \pab(\p_1)-\la\pab(\p_1)\ra_\ee \;.          \eqno(5.4)$$
Our problem is to understand the double-sum term on the right-hand side
of \eq(5.3).

For a fixed value of $\p_1$, each $\bigl|\tpab(\p_1)\bigr|$ is given roughly by
its RMS value in the eigenstate ensemble, which is $\D\pab(\p_1)$ as given by
\eq(3.13).  We have already seen that $\D\paa(\p_1)$ is extremely small, and
$\D\pab(\p_1)$ is not going to be any bigger when $\a\ne\b$.  In fact, using
the gaussian approximation of \eq(3.14) in \eq(3.13), we find
$$\bigl[\D\pab(\p_1)\bigr]^2 \simeq \bigl[\D\paa(\p_1)\bigr]^2\,
       \exp\bigl[-m(U_\a-U_\b)^2L^2/8\pi\hbar^2 U_\a\bigr]\;.    \eqno(5.5)$$
We see that we will have $\D\pab(\p_1)\ll\D\paa(\p_1)$ if $|U_\a-U_\b|/U_\a$
is much bigger than $(\hbar^2/mU_\a L^2)^{1/2}\sim \lam_\a/N^{1/2}L$,
a very small number.  [Note, though, that the precisely gaussian
form of the falloff is an artifact of \eq(3.14).]
For simplicity, let us assume that
$\D/\Ubar\ltwid\lambar/N^{1/2}L$, where $\lambar=(2\pi\hbar^2/mk\Tbar)^{1/2}$.
Then for the range of
$\a$ and $\b$ of interest, each $\D\pab(\p_1)$
is given by the right-hand-side of \eq(3.15) with $U_\a$ replaced by $\Ubar$.
We will need only the crudest approximations here, and so we write
$$\bigl|\tpab(\p_1)\bigr| \sim \D\pab(\p_1)
       \sim \bigl(L/\lambar\bigr)^{-3N/2}\;.                     \eqno(5.6)$$
However, we expect that the phase of $\tpab(\p_1)$ varies
wildly with $\a$ and $\b$.

Let $\Nc$ be the number of nonnegligible $C_\a$'s which appear in \eq(5.3);
$\Nc$ can be defined precisely via $\Nc^{-1}=\sum_\a|C_\a|^4$, and crudely
estimated as $\Nc\sim\nbar\D$, where $\nbar$ is the energy level density near
$U=\Ubar$ [cf.~\eq(4.6)].  The order of magnitude of each
nonnegligible $|C_\a|$ is then $\Nc^{-1/2}$ (so that $\sum_\a|C_\a|^2=1$).

Now consider doing the double sum in \eq(5.3).  If the phases of the $C_\a$'s
are not carefully correlated with those of the $\tpab(\p_1)$'s, then each of
the two sums will yield a random-walk result:  the square root of the number of
``steps'', $\Nc^{1/2}$, times the size of each step,
$C_\a\sim\Nc^{-1/2}$.  With an overall factor of
$\bigl(L/\lambar\bigr)^{-3N/2}$ from \eq(5.6), we get
$$\sum_{\a\b}C^*_\a C_\b\,e^{i(U_\a-U_\b)t/\hbar}\,\tpab(\p_1)
            \sim \bigl(L/\lambar\bigr)^{-3N/2}\;,                 \eqno(5.7)$$
which is again extremely small.  If we let
$\D/\Ubar$ be bigger than $\lambar/N^{1/2}L$, then the result is even smaller.

On the other hand, we can certainly set up an initial state which
is very far from thermal.  For example, we could give each particle the same
initial value (to within quantum uncertainties) for its individual momentum,
leading to an $\fqm(\p_1,0)$ which is sharply peaked at that value.
In this case, though, the phases of the $C_\a$'s must be correlated with
those of the $\tpab(\p_1)$'s in exactly the right way to produce the desired
nonthermal distribution $\fqm(\p_1,0)$.  In this case,
we want to see what happens as time evolves.

Let us begin with \eq(5.3) at $t=0$, with the phases of the $C_\a$'s carefully
chosen to give us a nonthermal distribution.  Now let the clock run.
Each of the off-diagonal ($\a\ne\b$) terms in the double sum begins
acquiring an extra phase; there are roughly $\Nc^2$ off-diagonal terms
in all.  The growing phase of each
individual term will cause its contribution to the sum to have a random
orientation in the complex plane once $|U_\a-U_\b|t/\hbar\gtwid 2\pi$.  We will
say that such a term has {\it decohered}.
The first terms to decohere (those with the largest difference between
$U_\a$ and $U_\b$) do so at a time $t\sim h/\D$.
The fraction of terms which have
decohered at later times is given roughly by $(\D-h/t)^2/\D^2$.
Thus the fraction of still coherent terms
at this time is roughly $1-(\D-h/t)^2/\D^2 \sim(h/\D)/t$ for $t\gg h/\D$.

Now, each of the coherent terms should give its usual contribution to the sum,
since its phase is still (almost) properly aligned, but the net contribution
of all the coherent terms will be suppressed by a factor of
$O(h/\D)t^{-1}$ due to their reduced population.
On the other hand, the terms which have decohered will contribute with
random phases.  Since almost all terms have decohered for $t\gg h/\D$,
their total contribution will be given by \eq(5.6), and is negligibly small.
Thus, overall we expect that {\it any nonthermal features present in the
initial distribution $\fqm(\p_1,0)$ will decay away with time
like $O(h/\D)t^{-1}$}.

We can test this conclusion with a very simple example.
The system we will analyze is classically integrable,
and so chaos plays no role in the following discussion.

Consider a single particle with mass $m=100$ in a two-dimensional circular box
with radius $R=1$; we also set $\hbar=1$.
The initial wave function for the particle is taken to be
$$\psi(\x,0) = \pi^{-1/2}a^{-1}\exp\bigl(i\p_0\cd\x\bigr)
               \exp\bigl(-\x^2\!/2a^2\bigr)\;.                    \eqno(5.8)$$
This is a gaussian wave packet of width $a$ at the center of the box, moving
with momentum $\p_0$.  If we Fourier transform into momentum space, we get
$$\tpsi(\p,0) = \pi^{-1/2}a\,
       \exp\bigl[-\half a^2(\p-\p_0)^2\,\bigr]\;.                 \eqno(5.9)$$
We will take $a=0.1$ and $p_0=100$.  Classically, the particle has energy
$E=p_0^2/2m= 50$, and just bounces back and forth, changing directions at times
$t=1,3,5,\ldots\,$.  Quantum mechanically, the uncertainty in the particle's
energy is $\D=p_0/\sqrt{2}ma \simeq 7$.  We can solve this problem exactly by
expanding $\tpsi(\p,0)$ in the energy eigenstate basis and using \eq(3.4).
Then we can compute the probability density for the particle to have its
initial momentum at time $t$; that is, we compute
$\bigl|\tpsi(\p_0,t)\bigr|{}^2$.  Let us see what we should expect for this
quantity, based on the general arguments outlined above.

First of all, note that the infinite time average of
$\bigl|\tpsi(\p_0,t)\bigr|{}^2$ is given by
$$\lim_{\tau\to\infty}{1\over\tau}\int_0^\tau dt\;\bigl|\tpsi(\p_0,t)\bigr|{}^2
  = \sum_\a\bigl|C_\a\bigr|^2 \,\bigl|\tpsi_\a(\p_0)\bigr|{}^2\;, \eqno(5.10)$$
where $\tpsi_\a(\p_0)$ is an energy eigenfunction in momentum space.  We expect
$\bigl|\tpsi(\p_0,t)\bigr|{}^2$ to approach its infinite time average at late
times, with late-time fluctuations of the same order of magnitude.  If
$\bigl|\tpsi(\p_0,t)\bigr|{}^2$ does not start out close to its infinite time
average at $t=0$, then it should decay towards that value like $O(h/\D)t^{-1}$.

The exact result for
$\bigl|\tpsi(\p_0,t)\bigr|{}^2$, normalized to its value at $t=0$, is shown
in \fig1.  We clearly see the classical bounces, as the probability to have
the initial momentum drops to zero at the first turning point, $t=1$, then
returns at $t=3$, etc.  However, the quantum probability does not return to its
initial value, but follows the $(h/\D)/t$ envelope predicted by the general
argument outlined above.  It finally drops down to its infinite time average,
with fluctuations of the expected size.  Thus we see that the simple phase
decoherence argument works very well for this example.

If we now compare the right-hand sides of \eqs(5.1) and \eq(5.10), we see the
analogy between the infinite time average of $\bigl|\tpsi(\p_0,t)\bigr|{}^2$
and the eigenstate ensemble average of $\fqm(\p_1,t)$.  The main difference is
that, because we have integrated out almost all the degrees of freedom,
the fluctuations of $\fqm(\p_1,t)$ about its average value are very small.

There is more we can learn from our simple example, however.  First, we have
seen that the classical motion is reflected in the quantum probabilities,
a fact which is expected to be true for classically chaotic systems
as well\r{longtime}.  This means that quantum initial states which can be
understood as representing classical initial conditions may thermalize
even faster, due to the effects of classical chaos.
If we can follow classical trajectories
(with initial quantum uncertainties) for some time, and classical chaos spreads
these out over a constant energy surface in phase space, then the system
has thermalized classically.  This argument may be needed in cases where the
initial distribution is so far from thermal that the quantum $O(h/\D)t^{-1}$
decay of its nonthermal features seems to take too long.  It also shows that
the $O(h/\D)t^{-1}$ rule need not be related to more traditional
diffusion times, which are more likely to reflect classical physics.

Second, consider \fig2, which shows the $|C_\a|$'s for this problem plotted
against the corresponding energy eigenvalues, $E_\a$.  While they form
a nice envelope, with mean energy $\bar E=50$ and uncertainty
$\D=7$, there is a great deal of fine structure.  This is needed
to get the very special initial state, localized at the origin and moving
in the $x$ direction at a particular speed.  We cannot, therefore, think
of replacing $C_\a$ by a smooth function of $E_\a$.  This is unfortunate,
since if we average \eq(4.4) over a smooth distribution of energy eigenvalues,
the contributions of the long periodic orbits are
suppressed\r{bog,berry89,berry91}, and the problem becomes much more tractable.
However, as we see in \fig2, such smoothing is physically far too
restrictive, since it would prevent us from considering a wide variety of
initial states which we could actually prepare in a real experiment.

Let us summarize the results of this section.  If we construct a particular
initial state for the hard-sphere gas at low density by superposing energy
eigenstates, each of which individually satisfies Berry's conjecture,
then we find that, at sufficiently late times, the quantum mechanical
prediction for the probability that any one particle has momentum~$\p$
is simply given by the Maxwell--Boltzmann distribution of \eq(1.1).
The probability that this will {\it not\/} be the case
is negligible, if we wait long enough.  Any nonthermal features of the
initial distribution for a single particle's momentum will decay away at
least as fast as $O(h/\D)t^{-1}$, where $\D$ is the quantum uncertainty in
the total energy.  Faster decays are possible, and likely if
the initial state has a classical interpretation.
Absolutely no averaging over initial states or times of measurement is
needed, in contrast with the classical case.

This concludes our analysis of the high energy, low density,
hard-sphere gas in the case that we
assume no symmetries of the wave function on exchange of individual
particles.  We will return to discuss lower energies and higher densities
in Sect.$\;$VII.  Now, though, we turn our attention to wave functions
which are either completely symmetric or completely antisymmetric
functions of the positions of the $N$ particles.

\vskip0.2in
\noindent\centerline{\bf VI.~~BOSONS AND FERMIONS}
\vskip0.2in

The detailed analysis in Sects.$\;$II and III required the assumptions of high
energy ($\lam_\a\ltwid a$) and low density ($Na^3\ll L^3$), which combine to
give $N\lam_\a^3\ll L^3$.  In Sect.$\;$II we noted that this is precisely the
condition needed for the \be and \fd distributions to be well approximated by
the \mb distribution.  Nevertheless, even though the corrections due to quantum
statistics may be numerically small, a valid formalism should be able to
reproduce them.  In this section we will see that the present formalism meets
this test.

Define a symmetrization operator $\Ppm$ via
$$\Ppm f(\ppp) = {1\over N!}\sum_{\rm perms}(\pm 1)^P \,
                            f(\p_{i_1},\ldots,\p_{i_N})\;,     \eqno(6.1)$$
where the sum is over the permutations of the indices, and $P$ is even (odd)
if the permutation is even (odd).  Now we can construct completely symmetric
and antisymmetric energy eigenfunctions analogous to those of \eq(2.2):
$$\psi^\pm_\a(\X) = \Na^\pm\int d^{3N}\!\!P\;A_\a(\P)\;
                    \d\bigl(\P^2-2m U_\a\bigr)\,
                    \Ppm\exp(i\P\cd\X/\hbar) \;.               \eqno(6.2)$$
We will need the generalization of \eq(2.9).  Using the same assumptions
as in Sect.$\;$III, namely $\lam_\a\ltwid a$ and $Na^3\ll L^3$, we get
$$\la\tpsi^{\pm\,*}_\a(\P)\tpsi^\pm_\b(\P')\ra_\ee
               = \d_{\a\b}\,\bigl(\Na^\pm\bigr)^2\,h^{3N}
                 \d(\P^2-2mU_\a)\,\Ppm\,\d^{3N}_D(\P-\P') \;.  \eqno(6.3)$$
We now want to compute
$$\la\paa^\pm(\p_1)\ra_\ee = \intp2 \la\tpsi^{\pm\,*}_\a(\P)
                                   \tpsi^\pm_\a(\P)\ra_\ee\;.  \eqno(6.4)$$
For nonsymmetric wave functions,
we found in Sect.$\;$III that $\la\paa(\p_1)\ra_\ee$ was equal to a \mb
distribution at a temperature $T_\a$ related to the energy eigenvalue $U_\a$
by the ideal gas formula $U_\a=\frac32 NkT_\a$.  We therefore expect
to find that $\la\paa^\pm(\p_1)\ra_\ee$ is given by a \be distribution
$\fbe(\p_1,T_\a)\equiv f^+(\p_1,T_\a)$ or a \fd distribution
$\ffd(\p_1,T_\a)\equiv f^-(\p_1,T_\a)$.
In statistical mechanics, these are usually computed using the grand canonical
ensemble, but in our case the number of particles is firmly fixed at $N$.
Thus we expect to find that $\la\paa^\pm(\p_1)\ra_\ee$ is equal to
$f^\pm(\p_1,T_\a)$ as given by the canonical, rather than the grand canonical,
ensemble.  Relevant formulae from the less familiar canonical
ensemble are gathered in the Appendix.

Let us warm up by computing $\Na^\pm$.  The normalization condition we need is
$$\intp1 \la\tpsi^{\pm\,*}_\a(\P)\tpsi^\pm_\a(\P)\ra_\ee = 1\;.  \eqno(6.5)$$
We must first evaluate
$$\Ppm\,\d^{3N}_D(\P-\P')\big|_{\P'=\P}
  = {1\over N!}\sum_{\rm perms}(\pm 1)^P \,
    \d^3_D(\p_1-\p_{i_1})\ldots\d^3_D(\p_N-\p_{i_N}) \;,  \eqno(6.6)$$
where $\d^3_D(\p)$ is assumed to satisfy the substitution rules of \eq(2.8)
with $N\to1$.  Now examine a particular term in this sum.
If a particular momentum is paired with itself,
we will say that it comprises a ``1-cluster.''
If a particular momentum is not paired with itself,
it will be part of an ``$m$-cluster'' of momenta which are all set equal to
each other by the (approximate) delta functions.
For each term, let $m_l$ be the number of momenta
in the $l$th cluster, with $m_1\le\dots\le m_C$.
Also let $C_m$ be the number of $m$-clusters,
and $C$ be the total number of clusters.
Obviously, we have the relations
$$\eqalign{
m_1+\ldots+m_C &= N\;,\cr
\noalign{\medskip}
C_1+\ldots+C_N &= C\;,\cr
\noalign{\medskip}
C_1+2C_2+\ldots+NC_N &= N\;.\cr}                           \eqno(6.7)$$
Each term in the sum in \eq(6.6) can now be labeled by a set of nondecreasing
integers $\{m\}\equiv\{m_1,\ldots,m_C\}$.
The number of terms with the same label is
$$A_{\{m\}}\prod_{l=1}^C (m_l-1)! \;,                             \eqno(6.8)$$
where $(m_l-1)!$ counts the number of ways momenta in the $l$th cluster
can be rearranged without breaking it into smaller clusters, and
$$A_{\{m\}}={N!\over(m_1 !\ldots m_C !)(C_1 !\ldots C_N !)}
                                                                  \eqno(6.9)$$
counts the number of inequivalent ways of assigning momenta to clusters.
Each cluster has one redundant delta function, which results in
a factor of $(L/h)^{3C}$.
Furthermore the $l$th cluster contributes a factor of $(\pm1)^{m_l-1}$
to $(\pm1)^P$.
Now from \eqs(6.3) and (6.5) we see that to determine $\Na^\pm$
we must multiply each term in \eq(6.6) by
$\d(\P^2-2mU_\a)$ and then integrate over all the momenta.
Under the integral,
when multiplied by a term labeled by $\{m\}$, we can make the replacement
$$\d(\P^2-2mU_\a) \to
  \d(m^\v_1\p_1^2+\ldots+m^\v_C\p_C^2-2mU_\a)\;,   \eqno(6.10)$$
All together, then, we have
$$\eqalignno{
\bigl(\Na^\pm\bigr)^{-2} &= h^{3N}
     \sum_{\{m\}} A_{\{m\}}\,(L/h)^{3C}
     \int d^3p_1\ldots d^3p_C \cr
     &\qquad \times
        \d(m^\v_1\p_1^2+\ldots+m^\v_C\p_C^2-2mU_\a)
        \prod_{l=1}^C (\pm1)^{m_l-1}(m_l-1)! \;.                  &(6.11)\cr}$$
The sum is over all $\{m\}$ with fixed $N$.  We now make the
change of variable $\p_i\to m_{\smash{i}}^{-1/2}\k_i$, which yields
$$\bigl(\Na^\pm\bigr)^{-2} = h^{3N}
    \sum_{\{m\}}A_{\{m\}}\,(L/h)^{3C}\,I_{3C}(2mU_\a)
    \prod_{l=1}^C (\pm1)^{m_l-1}(m_l-1)!\,m_{\smash{l}}^{-3/2}\;,\eqno(6.12)$$
where $I_D(x)$ is defined in \eq(3.9).  It turns out that
terms with $C\gg1$ dominate, and so we can use the large-$C$ formula
$$h^{3N}(L/h)^{3C}\,I_{3C}(2mU_\a)
     \simeq \lam_\a^{3N}(L/\lam_\a)^{3C}\,I_{3N}(2mU_\a)\;,      \eqno(6.13)$$
where $\lam_\a=(2\pi\hbar^2/mkT_\a)^{1/2}$, to rewrite \eq(6.12) as
$$\bigl(\Na^\pm\bigr)^{-2} =
  \lam_\a^{3N}\,I_{3N}(2mU_\a)\sum_{\{m\}}A_{\{m\}}
  \prod_{l=1}^C (L/\lam_\a)^3\,(\pm1)^{m_l-1}(m_l-1)!\,m_{\smash{l}}^{-3/2} \;.
                                                                 \eqno(6.14)$$
Now we can apply the Mayer cluster-expansion theorem\r{wannier},
which can be written as
$$\sum_{\{m\}}A_{\{m\}}\prod_{l=1}^C W_{m_l} =
  {\partial^N\over\partial z^N}\biggl\{
   \exp\biggl[\;\sum_{m=1}^\infty{z^m\over m!}\,W_m\biggr]\biggr\}
                                                          \bigg|_{z=0}\;,
                                                                 \eqno(6.15)$$
where, in our case,
$$W_m = (L/\lam_\a)^3\,(\pm 1)^{m-1}(m-1)!\,m^{-3/2}\;.          \eqno(6.16)$$
{}From \eq(A.3) of the Appendix, we have
$$\sum_{m=1}^\infty{z^m\over m!}\,W_m
                               = (L/\lam_\a)^3\,g^\pm_{5/2}(z)\;.\eqno(6.17)$$
So putting all of this together, we find
$$\eqalignno{
\bigl(\Na^\pm\bigr)^{-2} &=
      \lam_\a^{3N}\,I_{3N}(2mU_\a)\,
        {\partial^N\over\partial z^N}
        \biggl\{\exp\Bigl[\bigl(L/\lam_\a\bigr)^3\,
                g^\pm_{5/2}(z)\Bigr]\biggr\}\bigg|_{z=0}  \cr
\noalign{\medskip}
      &= \lam_\a^{3N}\,I_{3N}(2mU_\a)\,N!\,\zc\;,                 &(6.18)\cr}$$
where $\zc$ is the canonical partition function for a gas of noninteracting
bosons ($+$) or fermions ($-$) at a temperature $T_\a$
in a box of volume $L^3$ [cf.~\eq(A.8)].

Clearly we are on the right track!  Now we have to do it all over again, this
time leaving one of the $N$ momenta unintegrated.

Following the same logic which led us to \eq(6.12), we get
$$\eqalignno{
\la\paa^\pm(\p_1)\ra_\ee &=
\bigl(\Na^\pm\bigr)^2\,h^{3N}
     \sum_{\{m\}} A_{\{m\}}\,(L/h)^{3C}\,
     \sum_{i=1}^C(m_i/N)\,I_{3C-3}(2mU_\a-m_i\p_1^2) \cr
\noalign{\medskip}
&\qquad\times
   (\pm1)^{m_i-1}(m_i-1)!
   \prod_{l\ne i}(\pm1)^{m_l-1}(m_l-1)!\,m_{\smash{l}}^{-3/2}\;. &(6.19)\cr}$$
The differences from \eq(6.12) arise as follows.  First, we must choose which
cluster contains the unintegrated momentum $\p_1$; this gives the sum over
$i=1$ to $C$.  Then we must choose which of the $m_i$ momenta in the $i$th
cluster is unintegrated; this gives the factor of $m_i$.  Now we have
overcounted by $N$, which results in the factor of $1/N$.  The change to the
subscript and argument of $I$ results from not integrating $\p_1$, and the
factor of $m_{\smash{i}}^{-3/2}$ is missing because
we did not have to rescale $\p_1$.

Again, terms with $C\gg1$ dominate, and so we have
$$I_{3C-3}(2mU_\a-m_i\p_1^2) \simeq
(2\pi mkT_\a)^{-3/2}\exp\bigl(-m_i\p_1^2/2mkT_\a\bigr)\,I_{3C}(2mU_\a) \;.
                                                                 \eqno(6.20)$$
Then using \eq(6.13), we get
$$\la\paa^\pm(\p_1)\ra_\ee =
\bigl(\Na^\pm\bigr)^2\,\lam_\a^{3N}\,I_{3N}(2mU_\a)
     \sum_{\{m\}} A_{\{m\}}\sum_{i=1}^C V_{m_i}
      \prod_{l\ne i}W_{m_l}    \;,                               \eqno(6.21)$$
where $W_m$ is given by \eq(6.16), and
$$V_k = (L^3/N h^3)\,(\pm 1)^{k-1}k!\,
        \bigl[\exp\bigl(-\p_1^2/2mT_\a\bigr)\bigr]{}^k\;.        \eqno(6.22)$$
Starting with \eq(6.15), it is not hard to prove a generalization of it
which reads
$$\sum_{\{m\}}A_{\{m\}}\sum_{i=1}^C V_{m_i}\prod_{l\ne i}W_{m_l} =
  {\partial^N\over\partial z^N}\biggl\{
   \biggl[\;\sum_{k=1}^\infty{z^k\over k!}\,V_m\biggr]
   \exp\biggl[\;\sum_{m=1}^\infty{z^m\over m!}\,W_m\biggr]\biggr\}
                                                          \bigg|_{z=0}\;.
                                                                 \eqno(6.23)$$
In the present case, we have
$$\sum_{k=1}^\infty{z^k\over k!}\,V_m =
           {L^3\over Nh^3}\;{z\over\exp\bigl(\p_1^2/2mT_\a\bigr)\mp z}\;.
                                                                 \eqno(6.24)$$
Combining \eqs(6.17), (6.18), (6.21), (6.23), and (6.24), we finally get
$$\eqalignno{
\la\paa^\pm(\p_1)\ra_\ee &=
{1\over\zc}\,{1\over N!}\,{\partial^N\over\partial z^N}
       \biggl\{\exp\Bigl[\bigl(L/\lam\bigr)^3\,g^\pm_{5/2}(z)\Bigr]\;
{L^3\over Nh^3}\;{z\over\exp\bigl(\p_1^2/2mT_\a\bigr)\mp z}\biggr\}
                                                              \bigg|_{z=0} \cr
\noalign{\medskip}
&= f^\pm(\p_1,T_\a)\;,                                           &(6.25)\cr}$$
where $f^\pm(\p_1,T_\a)$ is the \be ($+$) or \fd ($-$) distribution as
predicted by the canonical ensemble [cf.~\eq(A.12)].
As expected, then, symmetrization or antisymmetrization of the wave function
changes the statistics from \mb to \be or \fd.

For the last time, we must study the fluctuations of $\paa(\p_1)$ that are
predicted by the eigenstate ensemble.  The relevant object is
$\D\pab^\pm(\p_1)$, defined by the obvious replacements in \eq(3.12).
$\D\pab^\pm(\p_1)$ is then given by \eq(3.13) with ${\cal P}_\pm$ acting on
$\d^{3N}_D(\P-\P')$.  Explicit evaluation of $\D\pab^\pm(\p_1)$ is a
fearsome combinatoric problem, but luckily a simple variation of the general
argument presented after \eq(3.13) still applies, and can be used to show
that $\D\pab^\pm(\p_1)$ is very small compared to
$\la\paa^\pm(\p_1)\ra_\ee\,\la\pbb^\pm(\p_1)\ra_\ee$.
Therefore eigenstate thermalization still holds, and the previous analysis
(in Sect.$\;$V) of time evolution still applies.

\vskip0.2in
\noindent\centerline{\bf VII.~~ DISCUSSION AND SPECULATION}
\vskip0.2in

Let us begin with a brief recap of the central results.  Berry's conjecture, as
applied to a gas of $N$ hard spheres in a box, states that each energy
eigenfunction appears to be a superposition of plane waves with wavelength
fixed
by the energy eigenvalue, but with random phases and gaussian random
amplitudes.  It is expected to apply only to systems which are classically
chaotic, and has been found to be valid (with corrections that do not affect
our conclusions) in simple chaotic systems.  Given \bc for the hard-sphere gas,
we have discovered the phenomenon of {\it eigenstate thermalization}:  each
energy eigenstate predicts a thermal distribution for the momentum of each
constituent particle.  This distribution is Maxwell--Boltzmann, Bose--Einstein,
or Fermi--Dirac, depending on whether the energy eigenfunctions are
nonsymmetric, completely symmetric, or completely antisymmetric functions of
the $N$ particle positions.  Then, a superposition of energy eigenstates with a
small fractional uncertainty in the total energy will also appear to be
thermal, {\it unless\/} the amplitudes and phases of the superposition
coefficients are carefully selected to avoid thermal behavior.  If this is done
initially, then the usual phase changes produced by hamiltonian time evolution
destroy the needed coherence, and any nonthermal features disappear as
$O(h/\D)t^{-1}$, where $\D$ is the uncertainty in the total energy.  However,
classical effects which are reflected in the quantum theory can result in
faster thermalization.

All of the analysis in Sects.$\;$II, III, and~VI was done in the limits of low
density: $Na^3\ll L^3$, where $a$ is the radius of a hard sphere and $L^3$
is the volume of the box, and high
energy: $\lam_\a\ltwid a$, where $\lam_\a=(2\pi\hbar^2/mkT_\a)^{1/2}$ is the
typical wavelength of one particle when the energy eigenvalue is
$U_\a=\frac32 NkT_\a$; numerically this means
$T_\a\gtwid(300/ma^2)\;$Kelvin, where $a$ is in angstroms and $m$ is in amu.
But what happens if we relax these constraints?

There are no fundamental difficulties with carrying out the analysis for
moderately higher densities.  All we need to do is use the exact formula
for the smeared delta function in momentum space, \eq(2.7).
In practice, though, this greatly complicates the calculations.
It would be very interesting to try to develop some sort of perturbative
(in $a/N^{1/3}L$) analysis, and compare the results with more standard
treatments of the hard-sphere Bose or Fermi gas\r{fetwal}.

Lower energies present an entirely different problem, since if we go low enough
in energy, \bc will break down.  The question is, how low can we go?  The
generic expectation is that \bc will be valid if the relevant wavelengths are
small enough to ``see'' the features which produce classical
chaos\r{mackauf88}.  For the hard-sphere gas, the relevant feature is the
nonzero radius~$a$ of each particle, which leads to $\lam_\a\ltwid a$.
However, this may not be good enough at high density\r{langer}.  Classically,
if the density is large enough to result in very slow diffusion of the
particles, then their positions will be correlated over long times; the
Lyapunov exponents are all very small.  We would then naturally expect that
these correlations are reflected in the quantum energy eigenfunctions, which
would mean that \bc is not valid.  In this case, a possible alternative
criterion for the validity of \bc is $\lam_\a\ltwid\ell$, where $\ell$ is the
classical mean free path of a particle, which can be much less than $a$.

Whatever the correct criterion turns out to be, at a low enough energy \bc will
break down, and we must ask what happens at lower energies.  One possibility is
that eigenstate thermalization will still be valid for a wide range of energy
eigenvalues, even though \bc is not.  The reason for this speculation appears
in the results of \eqs(3.15) and Sect.$\;$IV.  In \eq(3.15), we see that the
fluctuations about the mean, thermal value of $\paa(\p_1)$ in the eigenstate
ensemble are extremely small; experimentally, we can tolerate much larger
fluctuations.  Thus, we may also be able to tolerate significant violations of
Berry's conjecture without destroying eigenstate thermalization.
In Sect.$\;$IV, this speculation receives some more support.
Scars represent violations of \bc which are quite obvious when
one looks at the Wigner density of an energy eigenstate in phase space,
since there the scars appear with a ``signal-to-noise'' ratio of 1:1.
Once we integrate out all of the coordinates and most of the momenta,
however, the scars fade away almost completely.  The same should be
true of more generic violations of Berry's conjecture.  Thus, eigenstate
thermalization may still be valid at energies well below the threshold for the
validity of Berry's conjecture.

If we go even lower in energy, though, presumably eigenstate thermalization
will eventually cease to be valid.  If the system is this low in energy,
it will not be able to thermalize itself.  To find thermal behavior in
a system below its threshold for eigenstate thermalization,
we must couple it to an external
heat bath, such as the refrigeration apparatus in a low-temperature experiment.
Of course, once we have contact with a large, pre-existing heat bath,
all the usual results of statistical
mechanics can be applied without further worry.

The basic question we have been trying to address in this paper is
how such a heat bath might form in the first place.
We have seen that this will happen for a hard-sphere gas, provided that \bc is
satisfied by the energy eigenstates which are superposed to form the initial
state.  Whether or not other mechanisms exist for self-thermalization of
isolated quantum systems is an open question, one to which we hope to return.
Meanwhile we believe that the present results constitute a new foundation
for quantum statistical mechanics.  In particular, we have at least one
answer to the question of which quantum systems will approach thermal
equilibrium.  It is satisfying that this answer (those systems which obey
Berry's conjecture) is closely related to the answer from classical physics
(those systems which exhibit chaos).  In fact the situation in the quantum
theory is even better than it is in the classical theory, because we no longer
need to consider an ensemble of initial states.  Each and every superposition
of energy eigenstates obeying \bc will eventually yield a thermal distribution
for the momentum of a constituent particle, provided that we wait long
enough.  Absolutely no averaging of any kind is needed: not over initial
states, not over times of measurement, and not over hamiltonians.

Finally we would like to comment on the much-discussed question of an
appropriate definition for quantum chaos.  Some time ago, van Kampen suggested
that quantum chaos be defined as ``that property that causes a quantum system
to behave statistically''\r{vankam}.  If we replace ``behave statistically''
with ``obey the laws of statistical mechanics,'' then we have seen that the key
feature is Berry's conjectured properties of the energy eigenstates.  In
particular, properties of the energy eigenvalues (such as GOE rather than
Poisson statistics for the unfolded level spacings\r{bohigas}) have played no
role at all in the present work.  Steiner has suggested\r{steiner94} that
\bc be elevated to the status of the best definition of quantum chaos,
a proposal which we see to be equivalent to (our version of) van Kampen's.
More generally, in quantum mechanics, where time evolution is always linear
and therefore essentially trivial, the only place to encode the complexities
of the classical limit is in the energy eigenfunctions: that is where
quantum chaos, like thermal behavior, must be sought.

I am grateful to Walter Kohn, Jim Langer, Joe Polchinski, Andy Strominger,
and especially Doug Scalapino for helpful discussions.
I also thank Matthew Foulkes for pointing out an error in the original
version of this paper,
and the anonymous referee for several suggested improvements.
This work was supported in part by NSF Grant PHY--91--16964.

\vskip0.2in
\noindent\centerline{\bf APPENDIX.~~THE CANONICAL ENSEMBLE}
\noindent\centerline{\bf FOR BOSONS AND FERMIONS}
\vskip0.2in

We will use a notation close to that of\r{reichl}.  We ignore spin degrees
of freedom.

The canonical partition
function for $N$ noninteracting bosons or fermions in a box is given by
$$\zc = \prod_{l=0}^\infty\;\sum_{n_l=0}^{N_\pm}
              e^{-\b n_l E_l}\,\d_{N,n_1+n_2+\ldots}\;, \eqno(\rm A.1)$$
where $\b=1/kT$, $E_l$ is the $l$th energy eigenvalue for a single particle
in the box, and $N_+=\infty$ for bosons and $N_-=1$ for fermions.

Introducing the fugacity $z$, the grand canonical partition
function for $N$ noninteracting bosons or fermions in a box is given by
$$\eqalignno{
\zgc&= \prod_{l=0}^\infty\;\sum_{n_l=0}^{N_\pm}z^{n_l}e^{-\b n_l E_l} \cr
\noalign{\medskip}
    &= \exp\biggl[\mp\sum_{l=0}^\infty
                  \log\Bigl(1\mp z e^{-\b E_l}\Bigr)\biggr] \cr
\noalign{\medskip}
    &= \exp\biggl[\mp L^3 h^{-3} \int d^3p\,
                  \log\Bigl(1\mp z e^{-\b\p^2\!/2m}\Bigr)\biggr] \cr
\noalign{\medskip}
    &= \exp \Bigl[\bigl(L/\lam\bigr)^3\,g^\pm_{5/2}(z)\Bigr]\;.&(\rm A.2)\cr}$$
In the third line, we have replaced the sum over levels by an integral over
momenta (without separating out the zero mode, which would be necessary for
a discussion of Bose condensation).  In the fourth line, we have introduced the
thermal wavelength $\lam\equiv(2\pi\hbar^2/mkT)^{1/2}$, and the
Lerch transcendents
$$g^\pm_\nu(z)\equiv\sum_{m=1}^\infty{(\pm)^{m-1}z^m\over m^\nu}\;.
                                                               \eqno(\rm A.3)$$
$\zgc$ must be supplemented with the condition
$$\eqalignno{
N &= z\,{\partial\over\partial z}\log\zgc \cr
\noalign{\medskip}
  &= \bigl(L/\lam\bigr)^3\,g^\pm_{3/2}(z) \;,                 & (\rm A.4)\cr}$$
which can be thought of as fixing the value of $z$.  We will call the
positive real solution of this equation $z_0$.

The relation between $\zc$ and $\zgc$ can be found by writing the Kronecker
delta in \eq(A.1) as
$$\d_{N,n_1+n_2+\ldots} = {1\over 2\pi i}\oint dz\;z^{-N-1}\,
                                 z^{n_1+n_2+\ldots}\;,         \eqno(\rm A.5)$$
where the contour encloses the origin.  Substituting this into \eq(A.1) and
using the first and fourth lines of \eq(A.2) yields
$$\eqalignno{
\zc &= {1\over 2\pi i}\oint dz\;z^{-N-1} \zgc\cr
\noalign{\medskip}
    &= {1\over 2\pi i}\oint dz\;z^{-N-1}
       \exp \Bigl[\bigl(L/\lam\bigr)^3\,g^\pm_{5/2}(z)\Bigr]\;.&(\rm A.6)\cr}$$
Evaluating this integral approximately by stationary phase, treating both
$N$ and $\bigl(L/\lam\bigr)^3$ as large, results in
$$\zc =\left[2\pi\,\bigl(L/\lam\bigr)^3\,g_{1/2}^\pm(z_0)\right]^{-1/2}
                                                       \zgc\;, \eqno(\rm A.7)$$
where $\zgc$ is to be evaluated at $z=z_0$.
The fractional error in this approximate equality is of order $1/N$.
Note also that, using Cauchy's theorem, we can rewrite \eq(A.6) as
$$\zc = {1\over N!}\,{\partial^N\over\partial z^N}
        \biggl\{\exp\Bigl[\bigl(L/\lam\bigr)^3\,
                g^\pm_{5/2}(z)\Bigr]\biggr\}\bigg|_{z=0}\;.    \eqno(\rm A.8)$$

We would now like to compute the expected
fraction $\la f^\pm(\p)\ra d^3p$ of particles
with momentum in a range $d^3p$ around $\p$.
The expected fraction of particles with energy $E_l$ is given in either
formalism by
$$\eqalignno{
\la f^\pm_l\ra &= \la n^\pm_l\ra/N \cr
\noalign{\medskip}
               &= -\,{1\over N\b}\,{1\over Z^\pm}\,
                     {\partial Z^\pm\over\partial E_l}\;,     &(\rm A.9)\cr}$$
where $Z^\pm$ is either $\zc$ or $\zgc$.  Then converting to the normalization
required for continuous momenta gives
$\la f^\pm(\p)\ra = (L/h)^3\la f^\pm_l\ra$ with $E_l=\p^2/2m$.
In the grand canonical case this gives the well-known result
$$\la f^\pm(\p)\ra_{\rm GC} = {L^3\over Nh^3}\,
                        {z_0\over e^{\,\b\p^2\!/2m}\mp z_0}\;.\eqno(\rm A.10)$$
In the canonical case, this procedure gives
$$\la f^\pm(\p)\ra_{\rm C}={1\over\zc}\,{1\over 2\pi i}\oint dz\;z^{-N-1}\,
       \exp\Bigl[\bigl(L/\lam\bigr)^3\,g^\pm_{5/2}(z)\Bigr]\;
           {L^3\over Nh^3}\,{z\over e^{\,\b\p^2\!/2m}\mp z}\;.\eqno(\rm A.11)$$
Approximate evaluation of this integral by stationary phase gives
$\la f^\pm(\p)\ra_{\rm C} =\la f^\pm(\p)\ra_{\rm GC}$,
again with a fractional error of order $1/N$.
Also, we can again use Cauchy's theorem to write
$$\la f^\pm(\p)\ra_{\rm C} = {1\over\zc}\,
                                 {1\over N!}\,{\partial^N\over\partial z^N}
       \biggl\{\exp\Bigl[\bigl(L/\lam\bigr)^3\,g^\pm_{5/2}(z)\Bigr]\;
{L^3\over Nh^3}\,{z\over e^{\,\b\p^2\!/2m}\mp z}\biggr\}\bigg|_{z=0}\;.
                                                             \eqno(\rm A.12)$$

In the main text, we simplify the notation a bit via
$\la f^\pm(\p)\ra_{\rm C} \to f^\pm(\p,T)$.

\vfill\eject
\noindent\centerline{\bf REFERENCES}
\vskip0.2in
\references
\baselineskip=13pt

\refis{aurstein91}R. Aurich and F. Steiner, Physica D 48, 445 (1991).

\refis{aurstein93}R. Aurich and F. Steiner, Physica D 64, 185 (1993).

\refis{mackauf79}S. W. McDonald and A. N. Kaufman,
Phys. Rev. Lett. 42, 1189 (1979).

\refis{mackauf88}S. W. McDonald and A. N. Kaufman,
Phys. Rev. A 37, 3067 (1988).

\refis{shapiro88}M. Shapiro, J. Ronkin, and P. Brumer,
Chem. Phys. Lett. 148, 177 (1988).

\refis{sinai}Ya. G. Sinai, Ups. Mat. Nauk 25, 137 (1970)
[Russ. Math. Surv. 25, 137 (1970)].

\refis{berry77a}M. V. Berry, Phil. Trans. R. Soc. Lond. A 287, 237 (1977).

\refis{berry77b}M. V. Berry, J. Phys. A 10, 2083 (1977).

\refis{berry79}M. V. Berry, in {\sl Topics in Nonlinear Dynamics},
S. Jorna, ed. (AIP, New York, 1978).

\refis{berry81}M. V. Berry, in {\sl Les Houches XXXVI, Chaotic Behavior of
Deterministic Systems}, G.~Iooss, R. H. G. Helleman, and R. Stora, eds.
(North-Holland, Amsterdam, 1983).

\refis{berry89}M. V. Berry, Proc. R. Soc. Lond. A 423, 219 (1989).

\refis{berry91}M. V. Berry, in {\sl Les Houches LII,
Chaos and Quantum Physics}, M.-J. Giannoni, A.~Voros, and J. Zinn-Justin, eds.
(North--Holland, Amsterdam, 1991).

\refis{berrykeat}M. V. Berry and J. P. Keating,
Proc. R. Soc. Lond. A 437, 151 (1992);
O. Agam and S. Fishman, J. Phys. A 26, 2113 (1993).

\refis{bohigas}O. Bohigas, in {\sl Les Houches LII, Chaos and Quantum Physics},
M.-J. Giannoni, A.~Voros, and J. Zinn-Justin, eds. (North--Holland, Amsterdam,
1991).

\refis{gutz}M. C. Gutzwiller, {\sl Chaos in Classical and Quantum Mechanics}
(Springer--Verlag, New York, 1990).

\refis{KAM}A. N. Kolmogorov, Dokl. Acad. Nauk SSSR 98, 527 (1954);
V. I. Arnol'd, Usp. Mat. Nauk 18, 13 (1963) [Russ. Math. Surv. 18, 9 (1963)];
J. Moser, Nachr. Akad. Wiss. G\"ottingen 1 (1962).

\refis{krylov}N. S. Krylov, {\sl Works on the Foundations of Statistical
Mechanics} (Princeton Univ., Princeton, 1979).

\refis{zav}G. M. Zaslavsky, {\sl Chaos in Dynamic Systems}
(Harwood, Chur, 1985).

\refis{gaspard}P. Gaspard, in {\sl Quantum Chaos -- Quantum Measurement},
P. Cvitanovi\'c, I. Percival, and A. Wirzba, eds. (Kluwer, Dordrecht, 1992).

\refis{fetwal}A. L. Fetter and J. D. Walecka, {\sl Quantum Theory of
Many-Particle Systems} (McGraw-Hill, San Francisco, 1971).

\refis{vankam}N. G. van Kampen, in {\sl Chaotic Behavior in Quantum Systems},
G. Casati, ed. (Plenum, New York, 1985).

\refis{reichl}L. E. Reichl, {\sl A Modern Course in Statistical Physics}
(Univ. of Texas, Austin, 1980).

\refis{bog}E. B. Bogomolny, Physica D 31, 169 (1988).

\refis{heller84}E. J. Heller, Phys. Rev. Lett. 53, 1515 (1984).

\refis{heller91}E. J. Heller, in {\sl Les Houches LII,
Chaos and Quantum Physics}, M.-J. Giannoni, A.~Voros, and J. Zinn-Justin, eds.
(North--Holland, Amsterdam, 1991).

\refis{steiner94}F. Steiner, in {\sl Festschrift Universit\"at Hamburg 1994:
Schlaglichter der Forschung zum 75. Jahrestag}, R. Ansorge, ed.
(Dietrich Reimer Verlag, Hamburg, 1994).

\refis{wannier}G. H. Wannier, {\sl Statistical Physics}
(Wiley, New York, 1966).

\refis{jancel}R. Jancel, {\sl Foundations of Classical and Quantum Statistical
Mechanics} (Permagon, Oxford, 1969).

\refis{landau}L. D. Landau and E. M. Lifshitz, {\sl Statistical Physics}
(Permagon, Oxford, 1980).

\refis{longtime}S. Tomsovic and E. J. Heller,
Phys. Rev. Lett. 67, 664 (1991); Phys. Rev. E 47, 282 (1993);
M. A. Sep\'ulveda, S. Tomsovic and E. J. Heller,
Phys. Rev. Lett. 69, 402 (1992).

\refis{langer}J. S. Langer, private communication.

\refis{voros}A. Voros, Ann. Inst. H. Poincar\'e A 24, 31 (1976);
26, 343 (1977);
in {\sl Stochastic Behavior in Classical and Quantum Hamiltonian Systems\/},
G. Casati and J. Ford, eds. (Springer-Verlag, Berlin, 1979).

\refis{math}A. I. Shnirelman, Ups. Mat. Nauk 29, 181 (1974);
Y. Colin de Verdi\`ere, Comm. Math. Phys. 102, 497 (1985);
S. Zelditch, Duke Math J. 55, 919 (1987).

\endreferences
\vfill
\centerline{\bf FIGURE CAPTIONS}
\vskip0.2in
\item{Fig.$\;$1.} Solid line: $\bigl|\tpsi(\p_0,t)\bigr|{}^2\big/
                               \bigl|\tpsi(\p_0,0)\bigr|{}^2$
vs $t$ for a single particle in a two-dimensional circular box;
the initial wave function is a narrow gaussian at the center with
momentum $\p_0$.  Classically, the particle bounces off the wall at
$t=1,3,5,\ldots\,$.
Dashed line:  $(2\pi\hbar/\D)/t$, where $\D$ is the uncertainty in the
energy.  Dotted line: the infinite time average of the solid line.
\vskip0.2in
\item{Fig.$\;$2.} Expanding the wave function of \fig1 in energy
eigenstates yields expansion coefficients $C_\a$; here
$|C_\a|$ is plotted vs the energy eigenvalues $E_\a$.  There are 1736
energy eigenvalues in the plotted range, $20\le E\le 90$.
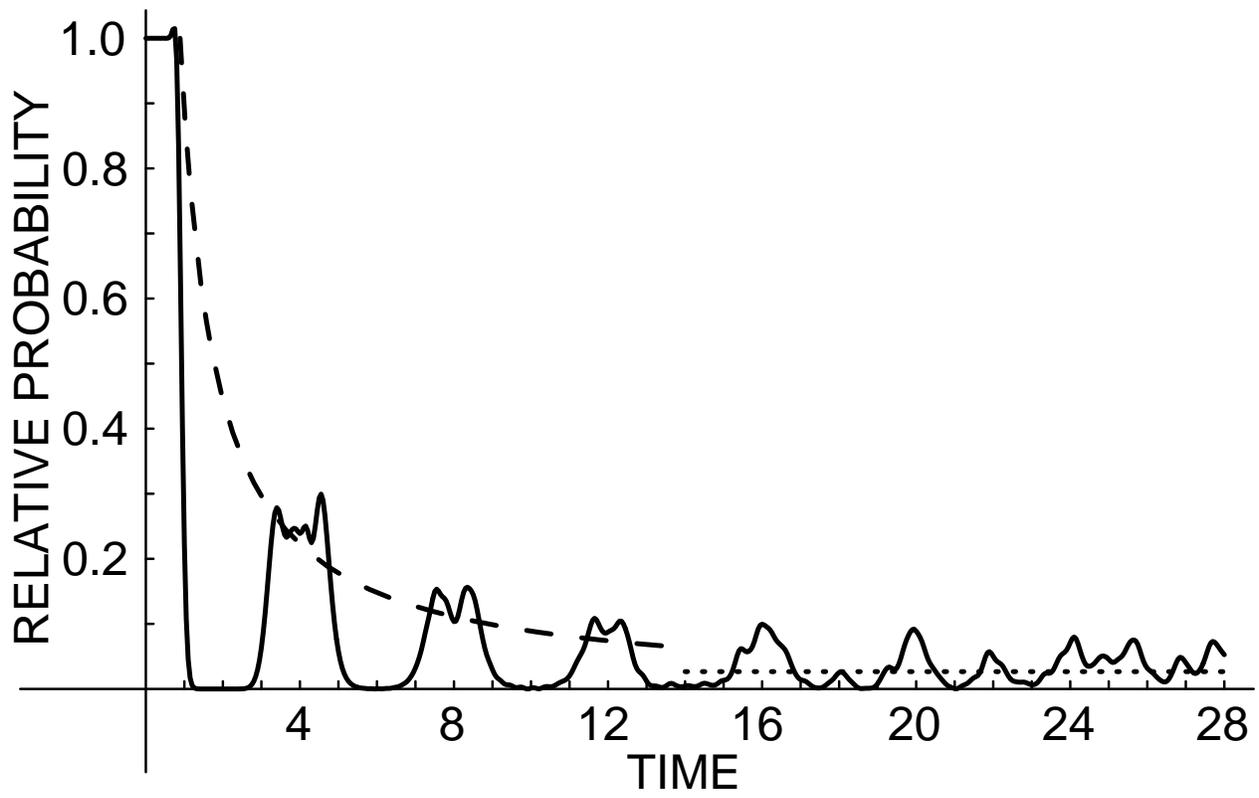
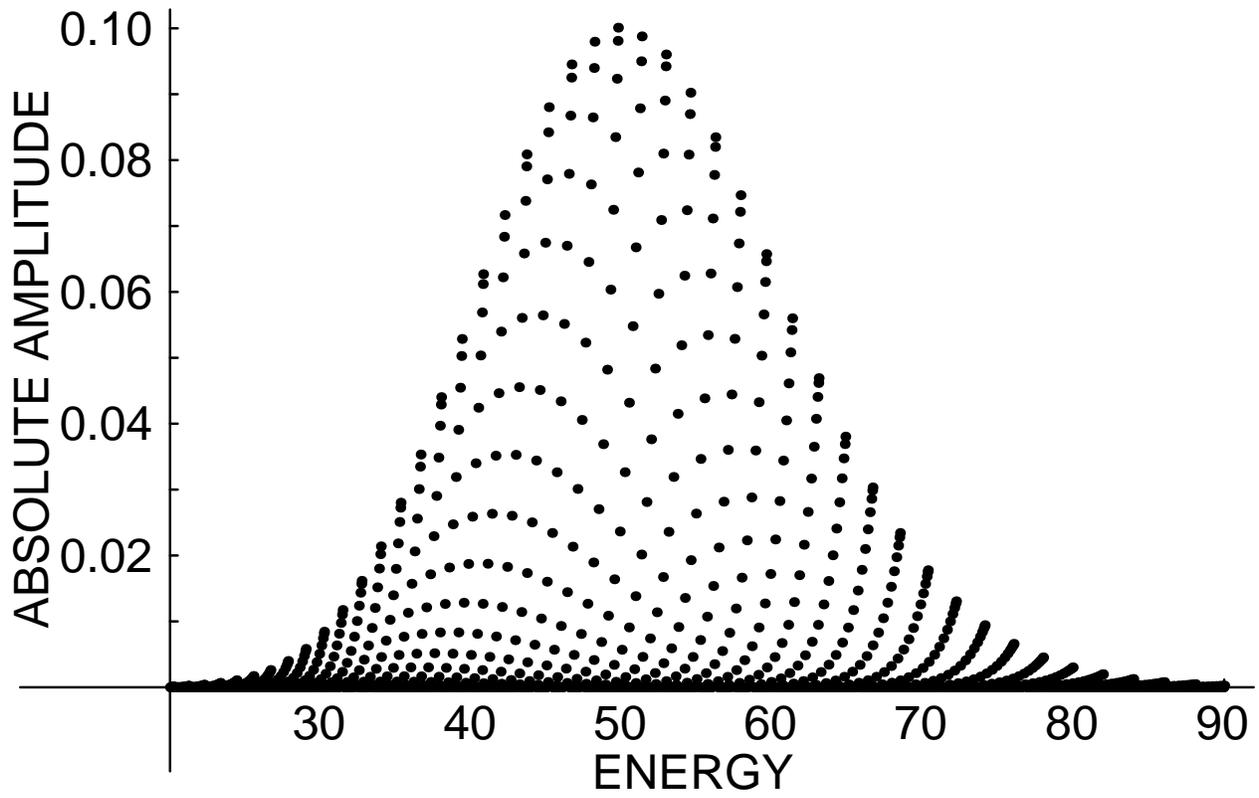
\end